\documentclass[sigconf,authorversion,nonacm]{acmart}
\AtBeginDocument{%
  }

\begin{document}

\title{Exploring Wikipedia Gender Diversity Over Time --- The Wikipedia Gender Dashboard (WGD)}

\author{Yahya Yunus}
\affiliation{%
  \institution{University of Queensland}
  \city{Brisbane}
  \country{Australia}
}
\email{s4705454@uq.edu.au}

\author{Tianwa Chen}
\affiliation{%
  \institution{University of Queensland}
  \city{Brisbane}
  \country{Australia}
}
\email{tianwa.chen@uq.edu.au}

\author{Gianluca Demartini}
\affiliation{%
  \institution{University of Queensland}
  \city{Brisbane}
  \country{Australia}
}
\email{g.demartini@uq.edu.au}

\renewcommand{\shortauthors}{Yunus et al.}

\begin{abstract}

The Wikipedia editors' community has been actively pursuing the intent of achieving gender equality. To that end, it is important to explore the historical evolution of underlying gender disparities in Wikipedia articles. This paper presents the Wikipedia Gender Dashboard (WGD), a tool designed to enable the interaction with gender distribution data, including the average age in every subclass of individuals (i.e. Astronauts, Politicians, etc.) over the years. Wikipedia APIs, DBpedia, and Wikidata endpoints were used to query the data to ensure persistent data collection. The WGD was then created with Microsoft Power BI before being embedded on a public website. The analysis of the data available in the WGD found that female articles only represent around 17\% of English Wikipedia, but it has been growing steadily over the last 20 years. Meanwhile, the average age across genders decreased over time. WGD also shows that most subclasses of `Person' are male-dominated. Wikipedia editors can make use of WGD to locate areas with marginalized genders in Wikipedia, and increase their efforts to produce more content providing coverage for those genders to achieve better gender equality in Wikipedia.

\end{abstract}

\settopmatter{printacmref=false}
\maketitle
\pagestyle{plain}

\section{Introduction}
Gender disparities have persisted for a long time because men are over-represented, even though they make up only half of the world’s total population. Many organizations are starting to take action to address the issue of gender inequality. The United Nations includes gender equality in one of their Sustainable Development Goals \cite{RN14}. 
%\cite{RN1}. 
The same challenge appears in  Wikipedia where the editors' community is proactively tackling it \cite{RN2}. Unfortunately, we are far from achieving gender equality \cite{RN14}. 

There are a few significant factors that cause gender imbalance in Wikipedia such as the authors' gender diversity \cite{RN3}.
About 80\% of Wikipedia editors are male. The narrow diversity among editors may play a part in the issue of gender imbalance as editors generally accommodate people from their group over other groups \cite{RN5}. It was also shown that Wikipedia editors usually tend not to work on under-represented entities \cite{RN6}.

Previous research has tried to address issues related to gender disparities \cite{RN7,RN8}. To name a few, \citet{RN7} discussed the issue of the gender gap in news by creating a dashboard to display the number of men and women mentioned by several news outlets in Canada over time. They found that men were mentioned three times more than women, and noted that authors' gender influences news coverage disparities.
Instead of news coverage, this work presents a dashboard that displays the gender distributions of Wikipedia articles under the class `Person'. \citet{RN8} estimated the completeness of Wikipedia using statistical estimators 
%from \cite{RN9}, 
finding a level of completeness close to 90\% across the genders in all the subclasses of  `Person' in Wikipedia articles. Subsequently, we analyze the gender distribution for all subclasses of `Person' such as genders for astronauts, monarchs, politicians, etc., assuming the data is almost complete.
On top of that, content bias in Wikipedia articles was explored across various language editions \cite{RN10, RN11}. They show that the English edition has the highest coverage \cite{RN11}, and Europeans and English-speaking culture  have more biographies compared to other languages \cite{RN10}. Thus, our work uses  exclusively English Wikipedia articles to maintain  simplicity and mitigate bias that might be posed when using articles from other language editions. The research by \citet{RN10} also shows that the female articles having a date of birth are more prominent than those having a date of death which might suggest a correlation between age and the number of female articles. Therefore, in our work we also include  age distribution data for each gender in each subclass of `Person'.
\citet{RN12} highlighted the herding effect \cite{RN15} and the effect of voters' heterogeneity \cite{RN12} as the editors are divided into two fractions called `inclusionists' and `deletionists'.
This paper aims to present a dashboard to facilitate agreement between them where the `inclusionists' work to include more articles of underrepresented gender and the `deletionists' work to delete articles of overrepresented gender in a subclass. 

Based on the limitations of previous research and prior work of Humaniki\footnote{Please see details from \url{https://humaniki.wmcloud.org/gender-by-dob}} on gender gap by year of birth and the assumptions made, the Wikipedia Gender Dashboard (WGD) has the following goals:
\begin{itemize}
\item Support Wikipedia editors in increasing their efforts to attain gender equality in Wikipedia articles by visualizing the gender distribution of Wikipedia articles for the class `Person' over time.
\item Provide high granularity of gender representation by showing gender disparities in every subclass of individuals for each of the years.
\item Calculate the mean age disposition of different genders in the subclasses of individuals articulated in Wikipedia to inform editors of underrepresented age cohorts related to the smaller gender group.
\end{itemize}

\section{System Design}
\subsection{Data Collection}
In this work we collected the instances of `Person' in Wikipedia, including their subclasses, gender, age, birth year, publication year, and Wikidata ID from the year 2001 to 2024. 
We used the following technical setup:
\begin{itemize}
    \item \textbf{SPARQLWrapper and Pandas Libraries:} Two key Python libraries named SPARQLWrapper and Pandas were implemented. SPARQLWrapper was used to query the DBpedia and Wikidata endpoints, while Pandas stored the collected data in a structured format\footnote{It is important to note that both DBpedia and Wikidata public endpoints impose limits, where DBpedia includes a 10,000-row restriction and a 40,000-row sorting cap for the query results, which was managed by leveraging offset and subquery techniques. Meanwhile, a Wikidata query is constrained to a 60-second runtime, which was avoided by querying one instance at a time for only the gender attribute.}.
\end{itemize}

\begin{itemize}
    \item \textbf{Querying Subclasses for Data Collection:} To maximize data retrieval, instances were queried by subclass (e.g., athletes, judges) on the DBpedia endpoint for all the attributes except for gender and publication year, which returned around 83\% of the total Person-related data on Wikipedia\footnote{Instances from the subclass ``Organization Member" were excluded due to redundancy with other categories.}.
\end{itemize}

\begin{itemize}
    \item \textbf{Gender and Publication Year Data:} Gender data was collected from Wikidata by querying the corresponding WikidataID for each instance which effectively overcame the 60-second query time limit\footnote{For publication year, the initial article creation date was extracted using the Wikipedia API.}.
\end{itemize}

\begin{itemize}
    \item \textbf{Error Handling and Data Export:} The Jupyter Notebook includes a robust retry mechanism for handling query errors, and the data was exported as CSV files that were combined during data preprocessing.
\end{itemize}

\begin{itemize}
    \item \textbf{Cloud Services for Persistent Data Collection:} The repeated querying of DBpedia and Wikidata SPARQL public endpoints took approximately 100 hours to collect all the data.
    Amazon SageMaker was used to enable long-running tasks and better integration with GitHub for version control and backups.
\end{itemize}

This data collection strategy lays the groundwork for analyzing gender disparities in Wikipedia and provides a reliable method for querying large public data endpoints.

\subsection{Data Preprocessing}
To ensure the data readiness for in-depth analysis, we conducted a series of steps to address data qualities issues related to duplicates, incorrect values, and missing values. 
A dataset consisting of 979,648 rows was created by combining all  the collected data. Seven attributes were included: subclass\footnote{subclass: a string representing the subclass assigned in Wikipedia of the `Person' instance (e.g., Judge, Monarch)}, instance\footnote{instance: a string denoting the name of the `Person' instance in Wikipedia.}, wikiDataID\footnote{wikiDataID: a string denoting the unique identifier of each instance in Wikipedia based on DBpedia. If not available, set to null.}, gender\footnote{gender: a string representing the gender assigned to the instance based on Wikidata. If unavailable, assigned to null.}, age\footnote{age: an integer indicating the person's age based on DBpedia. If exact age is not available, calculated as:
(a) age = death year – birth year. (b) If death year was unavailable (assumed to be alive), age = current year (2024) – birth year. (c) If both death and birth years are missing, set to null.}, birth year\footnote{birth year: a number from DBpedia. If unavailable, set to null.}, and publication year\footnote{publication year: a number denoting the first publication year of the instance retrieved via the Wikipedia API. Set to null if not available.}.

\subsubsection{Removing Duplicates}
The combination of instance name and wikiDataID had 134,353 duplicated rows that were removed. However, the individual columns of instance and wikiDataID still had duplicated values which should not be the case since wikiDataID is unique for each instance. To ensure the uniqueness we explored the possible cases as follows. (a) Multiple wikiDataIDs for the same instance. To resolve duplicated values in the instance column, instances were grouped by their name and subclass, revealing that 2625 instances in the dataset had multiple wikiDataID. The name of each wikiDataID for those instances was queried via DBpedia endpoint before identifying the correct wikiDataID by comparing the resulting name to the instance name in the dataset.
(b) Multiple instances with the same wikiDataID.
We found that 949 instances were sharing the same wikiDataID. To resolve duplicated values in the wikiDataID column, each of the 949 instance names (originally from the DBpedia endpoint) was cross-checked with the instance name on the Wikidata endpoint by querying the wikiDataID. However, it was not possible to verify correct matches between the instance names in DBpedia and Wikidata due to their naming inconsistencies. Thus, the wikiDataID for the 949 instances was set to null.

\subsubsection{Handling Incorrect Values}

We adopted different approaches to handle the incorrect values in the age and gender columns. We found that in age column, despite having no null values, the edge value includes the lowest age being -3395 and the highest age being 4431. To solve this issue, two edge cases were handled as follows. (a) For Age $\leq$ 0, the age of instances with age $\leq$ 0 that contain a unique WikidataID was cross-checked with the Wikidata endpoint, where instances that returned age >= 0 were updated in the dataset, while the age of instances that still returned negative values were set to null. Additionally, the age of the remaining instances without wikidataID having age $\leq$ 0 was set to null since there is no way to cross-check with Wikidata. (b) For Age $>$ Upper Limit,  an upper limit of 100 years was set, and 22,946 rows exceeded this limit. Out of those, the ages of all instances that returned a positive value were updated after cross-checking with the Wikidata endpoint, while the ages of others were stored as null. Finally, an additional Gaussian model was applied to detect outliers, setting all ages above 117 years to null. The gender column initially contains 52 unique values, and 28 of the genders, each representing one instance were removed because they were just hyperlinks that led to nowhere. This cleaning step resulted in 24 unique genders in the dataset.

\subsubsection{Handling Missing Values}
We addressed the missing values across each column as follows. For the gender column, we removed 47,838 rows with null values, opting not to use the existing `nomquamgender' package for binary gender classification for imputation as we aimed to maintain inclusivity for non-binary genders. Meanwhile, the 2,642 rows with null publication year were removed. Additionally, rows with missing age data were kept, considering their potential to provide insights related to demographics, publication year, subclasses, and gender.

\subsection{Dashboard Creation and Exploratory Data Analysis}

\begin{figure} 
    \centering
    \includegraphics[width=0.9\linewidth]{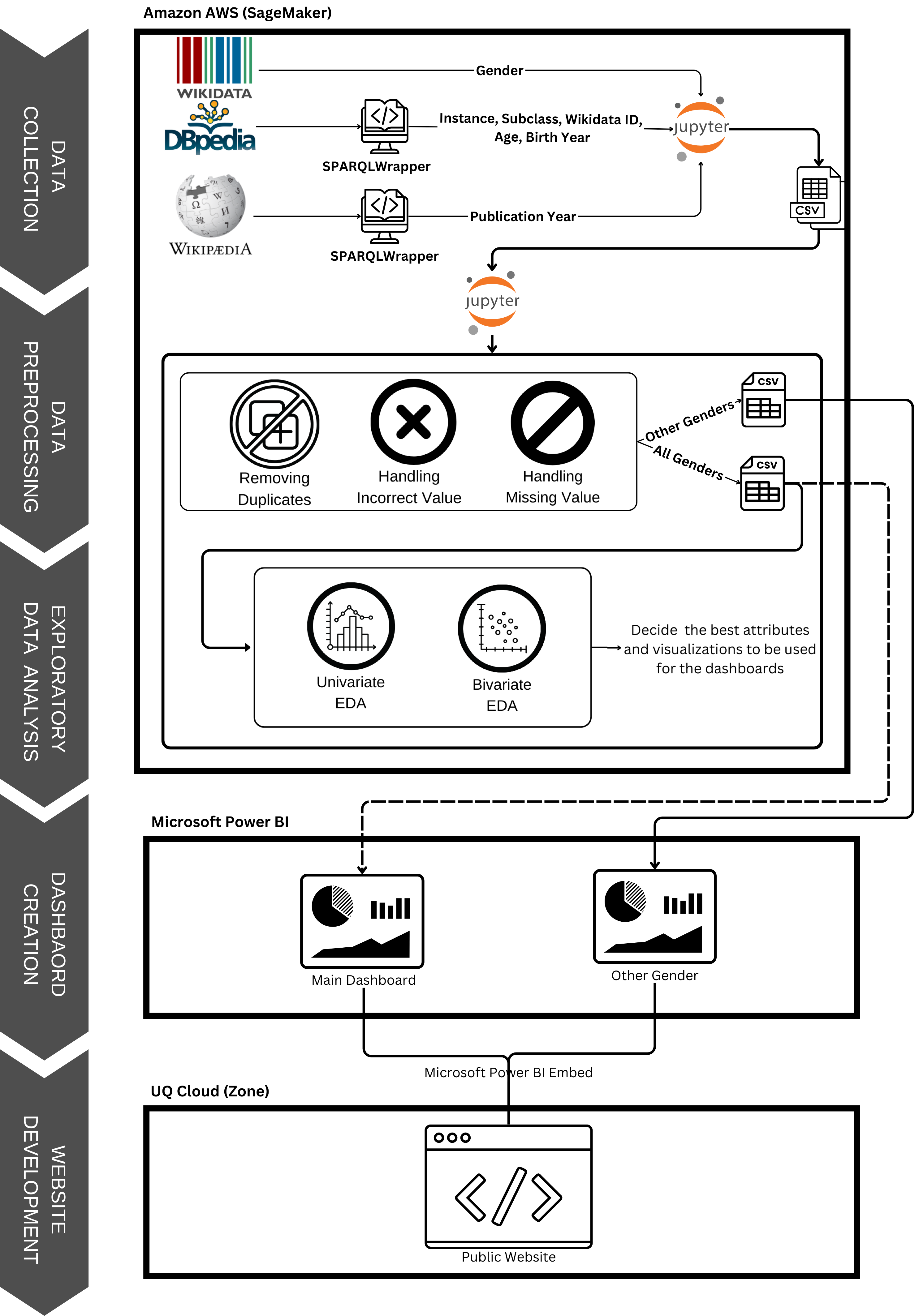}
    \caption{Top-down pipeline describing the system design.}
    \label{fig:Pipeline}
    \vspace{-4mm}
\end{figure}

The overall pipeline of the system design can be viewed in Figure~\ref{fig:Pipeline}. The Wikipedia Gender Dashboard was created via Microsoft Power BI using the preprocessed data where Gestalt principles \cite{RN18} were applied as they help to achieve a better signal-to-noise ratio by focusing on the information to be communicated (signal) and reducing clutter (noise) as much as possible. 
Two interactive dashboards were created where the main dashboard displays a histogram that depicts the total number of instances (articles) published over the years, a subclass filter, a pie chart representing the gender distribution, the number of articles, and the average age. Meanwhile, the other part of the dashboard which displays the same information using some different visualizations, focuses on genders that are not male and female as they are hard to analyze from the first dashboard being dominated by the binary genders. The demonstration of the dashboard can be viewed at the link\footnote{ Please see recorded demo at  \url{https://youtu.be/V_L6Q-S3MSA}}. The Exploratory Data Analysis (EDA) visualization techniques - Univariate and Bivariate EDA \cite{RN19} were used to find the best visualizations for the dashboards.

\subsection{Website Development}

A public website was developed for Wikipedia editors and general users to access the dashboards along with some project details. 
The dashboard was embedded in HTML into the website using iframe tags made available by Microsoft Power BI. 

\section{Demonstration}
\subsection{Comparative Analysis of Male and Female}
\begin{figure}
    \centering
    \includegraphics[width=1\linewidth]{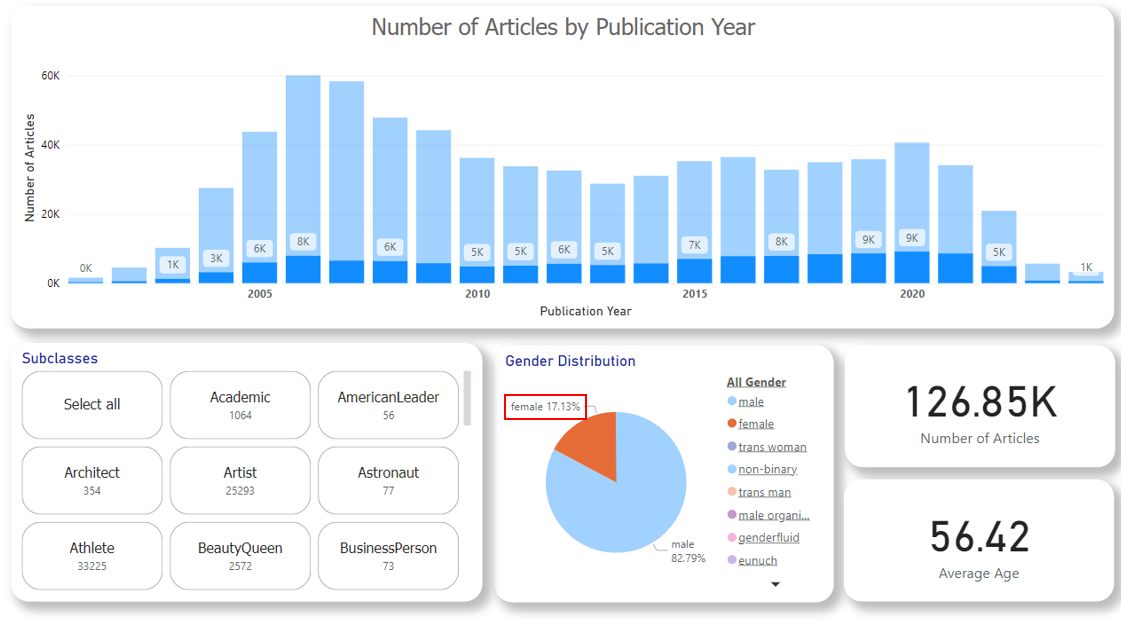}
    \caption{Main dashboard showing the distributions of female articles over the whole dataset.}
    \label{fig:MainDashboard}
    \vspace{-3mm}
\end{figure}

Between the years 2000 to 2022, females represent around 49.6\% of the global population on average \cite{RN17}. However, we find that females represent only 17.13\% (126,849) of Wikipedia as can be observed in Figure~\ref{fig:MainDashboard}.

\subsection{Gender and Average Age of Wikipedia Articles Published over the Years}
Through analysis, we find that the percentage of female articles published by Wikipedia increased steadily from 6.98\% in the year 2001 to 23.24\% in 2022. In addition, the average age across genders generally decreases over time, which could mean that more young `Person' articles have been published on Wikipedia in recent years.

\subsection{Gender and Average Age Grouped By Subclasses}
\begin{figure}
    \centering
    \includegraphics[width=0.9\linewidth]{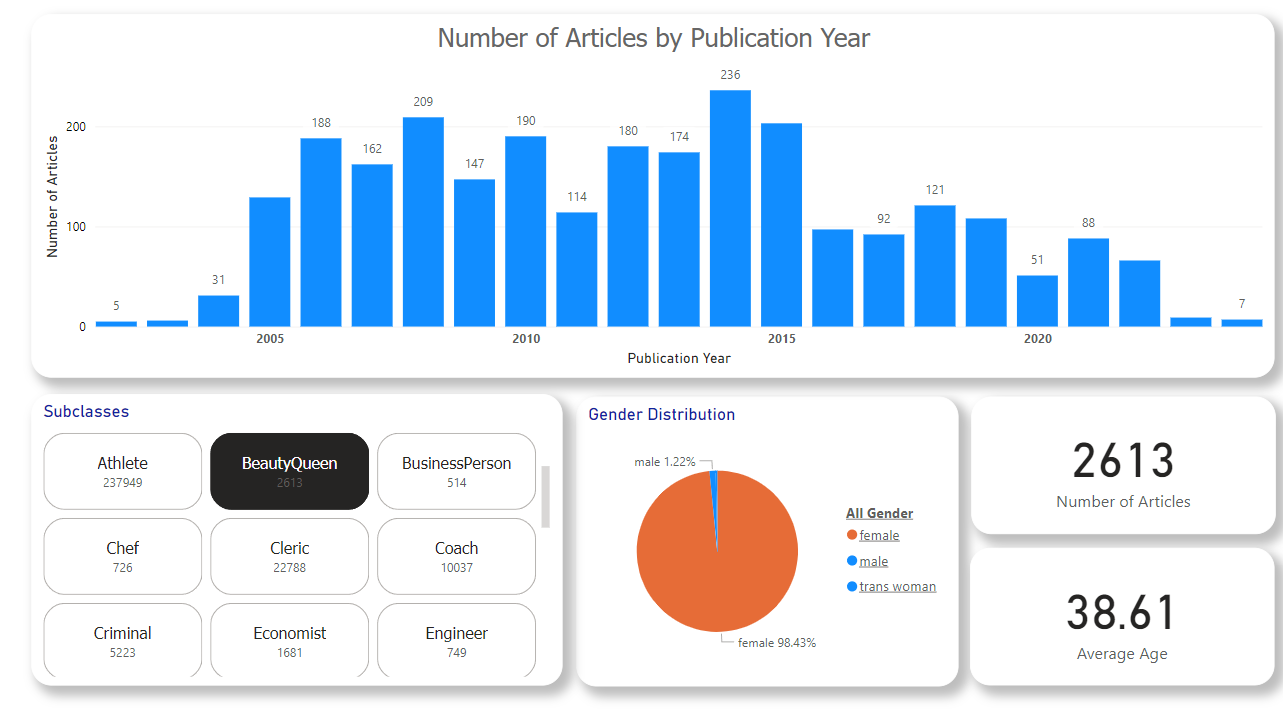}
    \caption{Distribution in the Beauty Queen subclass.}
    \label{fig:BeautyQueen}
\end{figure}

\begin{figure}
    \centering
    \includegraphics[width=0.9\linewidth]{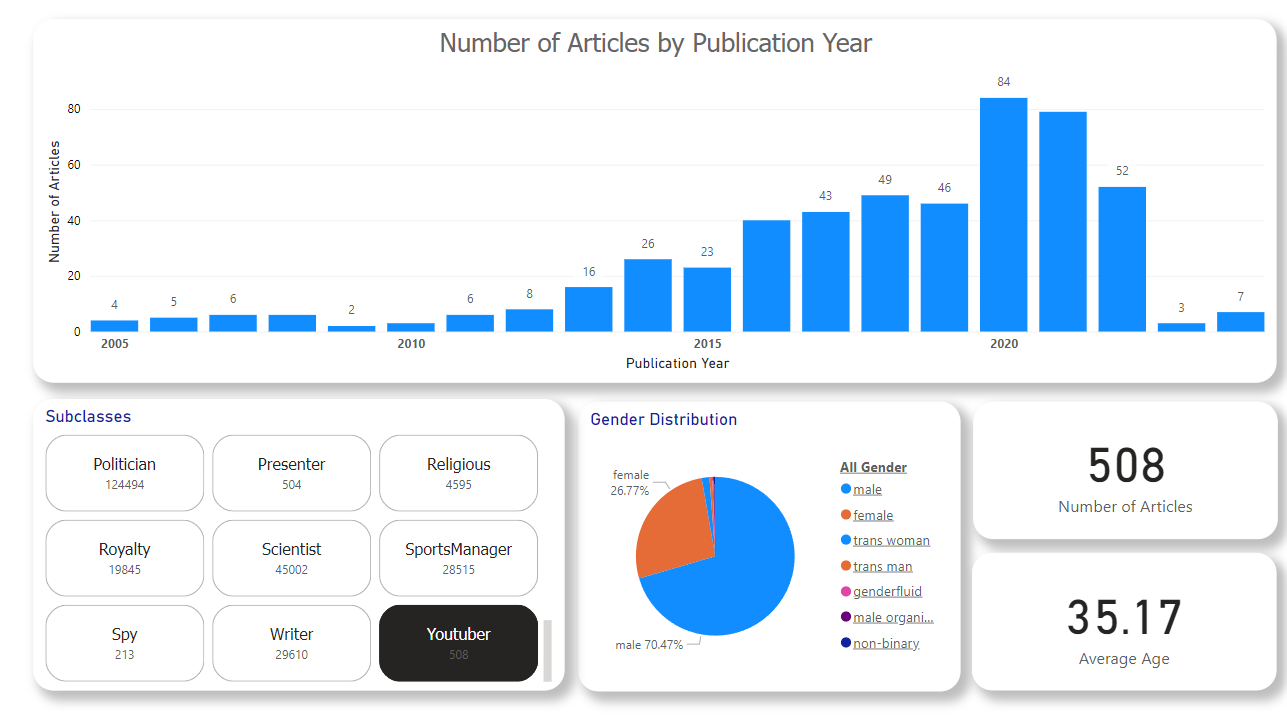}
    \caption{Distribution in the  Youtuber subclass.}
    \label{fig:Youtuber}
\end{figure}

Males dominate all the subclasses of `Person' in Wikipedia except for BeautyQueen (Figure~\ref{fig:BeautyQueen}), Model, and PlayboyPlaymate which are female-dominated. The average age of articles in most subclasses is quite high as only the BeautyQueen (Figure~\ref{fig:BeautyQueen}), Model, and Youtuber subclass (Figure~\ref{fig:Youtuber}) have an average age of less than 40 years.

\subsection{Other Genders}
\begin{figure}
    \centering
    \includegraphics[width=0.9\linewidth]{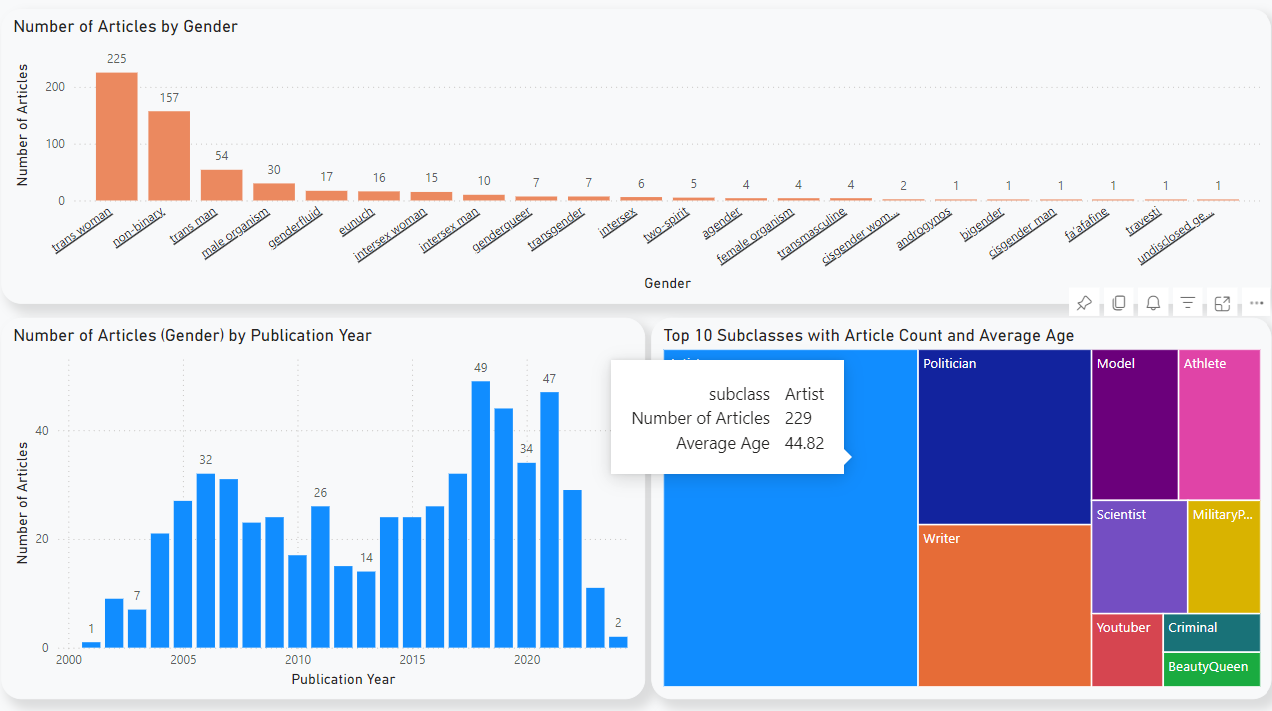}
    \caption{`Other Genders' dashboard that visualizes the instances having gender other than male or female.}
    \label{fig:OtherGenders}
    \vspace{-5mm}
\end{figure}

Figure~\ref{fig:OtherGenders} shows the `other genders' dashboard where it can be observed that trans-women have the most articles published (225 articles) on Wikipedia after males and females. Additionally, 2018 is the year with the most number of articles being published for other genders. Finally, the subclass with the most articles for genders that are not male or female (229 articles) is Artist with an average age of 44.82.

In this work, we developed the Wikipedia Gender Dashboard (WGD) to provide the Wikipedia editor community with insights on  gender and age distribution  within Wikipedia articles for different subclasses of `Person' from 2001 to 2024. Our system shows that while only 17\% of current Wikipedia articles feature females — up from 6.98\% in 2001 — the proportion of female articles has gradually increased, indicating the  impact of the efforts by the Wikipedia editors' community. Additionally, the average age across genders has decreased, suggesting that Wikipedia's standards may have evolved to include younger individuals. Despite these changes, most subclasses remain male-dominated.

Our work offers valuable contributions to the Wikipedia editors community, facilitating the identification and addressing the gaps in content related to marginalized genders, thereby supporting efforts to achieve gender equality in Wikipedia. Our work is not without limitations. First, our method is constrained by the assumptions made in age calculations. When the date of death is unavailable, we assumed the instances were alive which can inaccurately represent the age of deceased individuals with unrecorded death dates. Second, the generalizability of our findings might be limited by the collected data, which may not fully capture the diversity of gender categories globally. To enhance the robustness of our research, future work could focus on developing a machine learning classifier to more accurately label an instance to ``alive" or ``deceased". The age of an ``alive" instance could then be calculated as the difference between the current year and the birth year; otherwise, the age could be set to null, thus improving the precision of our dataset and validity of our conclusions.

In summary, we introduced a new dashboard that aims to empower the Wikipedia community to take data-driven decisions on how to manage gender balance on Wikipedia. For example, they may want to take the decision not to add new `male' persons to the sub-class Astronauts or to only add new articles for `female' persons to the subclass `scientist' for a certain period of time.

\textbf{Acknowledgments.} This work is partially supported by the ARC Training Centre for Information Resilience (Grant No. IC200100022) 
and by the Wikimedia Research Fund (Grant ID G-RS-2303-12081).

 \bibliographystyle{ACM-Reference-Format}

\bibliography{output}

\end{document}